\documentclass[12pt]{article}
\usepackage[margin=1in]{geometry}
\usepackage{amsmath,amssymb,amsfonts,bm,mathtools}
\usepackage{graphicx,booktabs,longtable,array,multirow,threeparttable}
\usepackage{caption,float,placeins}
\usepackage[backend=biber,style=numeric-comp,sorting=none,natbib=true]{biblatex}
\usepackage{setspace,microtype}
\usepackage{xcolor}
\usepackage[colorlinks=true,linkcolor=blue!45!black,citecolor=blue!45!black,urlcolor=blue!45!black]{hyperref}
\usepackage{authblk}
\usepackage{siunitx}
\usepackage{fancyhdr}
\newcommand{\Cor}{\operatorname{Cor}}

\title{\textbf{A Hybrid Spatial Statistical Learning Framework for Individualized Probability Estimation: Application to Multisite Autism Neuroimaging}}
\author[1]{Montserrat Fuentes}
\author[2]{Veronica B. Patterson}
\affil[1]{Department of Mathematics, St. Edward's University, Austin, Texas, USA}
\affil[2]{Department of Statistics, Rice University, Houston, Texas, USA}
\date{}

\begin{document}
\maketitle

\noindent\textbf{Correspondence:} V. Patterson, vp30@rice.edu, Department of Statistics, Rice University, Houston, Texas, USA.

\begin{abstract}

Autism spectrum disorder is a heterogeneous neurodevelopmental condition whose functional brain organization varies across individuals and imaging centers. Resting-state functional connectivity provides an opportunity to investigate this variation, but its analysis is complicated by high dimensionality, strong dependence among connections, and substantial differences in acquisition protocols and participant populations across sites. Existing approaches commonly rely on either detailed connectivity features or lower-dimensional network summaries and are often evaluated primarily through discrimination. A model may therefore rank participants reasonably well while producing probability estimates that are unreliable when transported to a new imaging environment.

We introduce a Hybrid Spatial Statistical Learning Framework for individualized probability estimation in multisite autism neuroimaging. The framework treats edge-level connectivity, graph-based network organization, and participant characteristics as complementary representations of a common probability target. Detailed connectivity retains fine-scale discriminatory information, while broader representations are used to stabilize the probability scale under site heterogeneity. All model development and preprocessing are performed within complete site-held-out validation, providing a direct assessment of transportability to unseen acquisition environments.

Simulation studies show that the hybrid preserves the discrimination of the strongest fine-scale model while improving probability accuracy as between-site heterogeneity increases. In an analysis of 860 participants from 20 Autism Brain Imaging Data Exchange sites, the hybrid retained the edge model's ROC AUC, increasing it only from \(0.646\) to \(0.647\), while reducing the Brier score from \(0.256\) to \(0.243\) and the log loss from \(0.733\) to \(0.684\). These represent reductions of \(4.9\%\) and \(6.6\%\), respectively. The hybrid reduced Brier-score error at 18 of the 20 held-out sites.

These findings demonstrate that multiscale integration can improve the reliability and transportability of individualized autism probabilities. More broadly, the framework provides an interpretable approach for probability estimation from complex, spatially dependent biomedical data collected across heterogeneous institutions.

\end{abstract}


\section{Introduction}
Autism spectrum disorder is a heterogeneous neurodevelopmental condition whose biological basis remains difficult to characterize \citep{Lord2020}. Its clinical presentation varies widely, and no single pattern of brain organization is expected to describe all individuals on the spectrum. Resting-state functional magnetic resonance imaging (rs-fMRI) offers a way to study this heterogeneity by measuring coordinated activity across distributed brain regions. The resulting functional-connectivity networks have become an important resource for investigating how large-scale brain organization differs in autism \citep{DiMartino2014ABIDE}. They also create a demanding statistical problem. The signal is distributed across many connections, those connections are strongly dependent, and the data are often collected at sites that differ in ways unrelated to autism.

The Autism Brain Imaging Data Exchange (ABIDE) makes this challenge especially clear. ABIDE brought together imaging and phenotypic data from many independent institutions and created one of the largest public resources for autism neuroimaging \citep{DiMartino2014ABIDE}. Its scientific value comes not only from the larger sample but also from the diversity of the contributing studies. That diversity is essential for understanding whether findings extend beyond a single center. It also means that scanner characteristics and recruitment practices can shape the observed connectivity patterns. A model may therefore perform well when the same sites appear in both training and testing, yet fail when it is moved to a new imaging environment.

Functional connectivity is difficult for another reason. A parcellation with \(R\) brain regions yields \(R(R-1)/2\) pairwise connections, so the number of candidate features can quickly exceed the number of participants. These features are not exchangeable. Connections that share a region are statistically related, and broader functional systems create dependence across distant parts of the brain. Treating the connectivity matrix as an ordinary feature vector can therefore discard important information about how the features are organized.

Several classes of methods have been used to address these challenges. Regularized regression and other machine-learning procedures can stabilize prediction in high dimensions \citep{Hastie2009,Abraham2014,Pedregosa2011}. Graph-based approaches represent the brain as a network and use global or regional summaries to describe its organization \citep{Nielsen2013,Heinsfeld2018,Parisot2018,Ktena2018}. Harmonization methods seek to reduce systematic differences across sites \citep{ComBat2007,Fortin2018}. Each of these approaches is valuable, but none resolves the full problem. Edge-based models retain detailed signal but can be sensitive to site-specific perturbations. Graph summaries are more compact, but they may remove the localized patterns that carry most of the discrimination. Harmonization can reduce measured site effects, yet it cannot ensure that a prediction model will remain reliable at a center that was not represented during training.

A further limitation is that most neuroimaging studies emphasize classification or ranking. Accuracy and the area under the receiver operating characteristic curve are useful measures, but they do not show whether the predicted probabilities are numerically trustworthy. A model can rank participants correctly and still assign probabilities that are too extreme. This distinction matters in multisite studies because a change in acquisition environment may alter the scale of the predictions even when the ranking is largely preserved. Proper scoring rules such as the Brier score and log loss evaluate the quality of the probability itself and therefore provide information that classification metrics cannot \citep{Brier1950,Dawid1982,GneitingRaftery2007,Steyerberg2010,VanCalster2019}. Site-held-out validation is also needed because random subject-level splits can place the same acquisition environment in both the training and test samples \citep{Varoquaux2018,Scheinost2019}.

These limitations point to a methodological gap. The available methods tend to favor one representation of the connectome at a time. Detailed edges and graph summaries are often treated as alternatives, even though they describe different scales of the same biological system. When they are combined, the combination is usually motivated by predictive performance rather than by the spatial structure of the data. What is needed is a framework that preserves the discrimination found in fine-scale connectivity while allowing broader network information to stabilize the resulting probabilities. That framework must also be evaluated under a genuine change of site rather than under a random split of participants.

Spatial statistics provides a natural foundation for this development. Work on nonstationary covariance modeling has shown that a single global dependence structure may obscure meaningful changes across space \citep{Fuentes2001,Fuentes2002Biometrika,Fuentes2005JASA}. Spectral and multivariate spatial methods extend this idea by representing complex dependence through components that operate at different scales \citep{TerresFuentes2018,FuentesReichLee2008}. Bayesian model combination provides a related principle: imperfect sources may be integrated when each offers a different view of a common latent quantity \citep{FuentesRaftery2005}. These ideas suggest that edge-level connectivity and graph-level organization should not be forced to compete. They can instead be treated as complementary projections of the same underlying disease signal.

Motivated by this perspective, we introduce a new Hybrid Spatial Statistical Learning Framework for individualized probability estimation in multisite autism neuroimaging. The framework uses detailed connectivity features to retain local and distributed signal. It uses graph summaries to represent broader network organization. A small participant-level component provides context for variation not captured by imaging alone. The component probabilities are estimated without access to the held-out sites and are combined through a prespecified convex rule. This estimator gives the edge model the primary role in discrimination while allowing the coarser representations to moderate probabilities that become unstable under site shift.

The principal methodological contribution is a multiscale statistical framework that combines edge-level connectivity, graph-level organization, and participant characteristics to improve the quality and transportability of individualized probabilities across imaging sites. The method does not simply average unrelated algorithms. Each component has a prespecified role tied to a particular spatial scale, and the combined estimator approximates a common probability target. The framework is designed around probability quality rather than classification alone. Its evaluation also treats acquisition site as part of the statistical problem: entire sites are excluded during validation, so performance reflects transportability to new imaging environments rather than interpolation among sites already represented during training.

This contribution is needed because no single representation performs all of these tasks well. A high-dimensional edge model can preserve disease-related detail but may become overconfident when the site distribution changes. A graph model can provide stability but may not discriminate autism from control participants on its own. The proposed framework uses the strengths of both without claiming that either is sufficient. The resulting method is computationally practical, aligned with the spatial organization of the connectome, and directly targeted to the quality of individualized autism probabilities.

We study the framework in two complementary settings. The simulation study isolates the roles of spatial dependence and site heterogeneity and shows when multiscale integration improves out-of-site probability estimation. The empirical analysis uses quality-controlled ABIDE rs-fMRI and evaluates the method across acquisition sites that were not involved in fitting the model. Together, these analyses assess both the statistical mechanism of the proposed estimator and its performance in a substantive biomedical application.

The remainder of the paper is organized as follows. Section~\ref{sec:data} describes the ABIDE resource, the analytic cohort, the preprocessing pipeline, and the construction of the functional-connectivity data. Section~\ref{sec:methods} develops the Hybrid Spatial Statistical Learning Framework and explains its multiscale interpretation, estimation procedure, and site-held-out validation design. Section~\ref{sec:simulation} presents the simulation study. Section~\ref{sec:application} reports the autism neuroimaging application and the resulting out-of-site probability performance. Sections~\ref{sec:discussion} and \ref{sec:conclusion} discuss the statistical and biomedical implications and summarize the main conclusions.

\section{Multisite autism neuroimaging data}
\label{sec:data}
Autism is a scientifically important setting for spatial statistical learning because its neurobiology is distributed and heterogeneous. Resting-state functional connectivity offers a noninvasive description of coordinated neural activity across the brain. At the same time, connectivity measurements are high dimensional, strongly dependent, and sensitive to acquisition conditions. The ABIDE consortium provides a large multisite resource in which these scientific and statistical features can be studied together.

ABIDE-I combines imaging and phenotypic data contributed by independent research institutions \citep{DiMartino2014ABIDE}. Its value is not limited to a larger sample size. The participating sites used different scanners, sequence parameters, recruitment strategies, and participant populations. The resulting diversity better represents the conditions under which multisite biomedical models must operate. It also creates the possibility that a prediction algorithm will learn center-specific features instead of transportable disease-related information. We therefore use acquisition site as a central element of the study design rather than as an incidental nuisance variable.

The analysis used quality-checked derivatives from the Preprocessed Connectomes Project \citep{Craddock2013PCP}. Images were processed with the Configurable Pipeline for the Analysis of Connectomes. Regional time series were obtained using the Automated Anatomical Labeling atlas \citep{TzourioMazoyer2002}. Temporal band-pass filtering was applied, and global-signal regression was not used. These choices define one reproducible preprocessing configuration rather than an assertion that a single pipeline is optimal for all neuroimaging questions.

The downloaded resource contained 871 participants with corresponding AAL regional time-series files. Every file was checked for dimensional consistency, finite values, sufficient temporal length, and variation within each region. Eleven participants were excluded because several regional signals were constant or nearly constant and could not support valid correlation estimates. The final analytic cohort contained 860 participants from 20 sites. There were 397 participants with autism and 463 controls. The mean age was 16.83 years with a standard deviation of 7.47 years. The sample included 141 female participants, representing 16.4\% of the cohort. Mean framewise displacement was 0.109 with a standard deviation of 0.116. The median number of time points was 176, with a range from 78 to 296. Table~\ref{tab:cohort} summarizes the demographic, diagnostic, motion, and time-series characteristics of the final analytic cohort.

\begin{table}[H]
\centering
\caption{Characteristics of the analytic ABIDE cohort.}
\label{tab:cohort}
\begin{threeparttable}
\begin{tabular}{lr}
\toprule
Characteristic & Value\\
\midrule
Participants & 860\\
Autism spectrum disorder & 397\\
Controls & 463\\
Acquisition sites & 20\\
Age, mean (SD), years & 16.83 (7.47)\\
Female, $n$ (\%) & 141 (16.4)\\
Mean framewise displacement, mean (SD) & 0.109 (0.116)\\
Time points, median (range) & 176 (78--296)\\
\bottomrule
\end{tabular}
\begin{tablenotes}\small
\item The table describes the 860 participants retained after regional time-series quality control.
\end{tablenotes}
\end{threeparttable}
\end{table}

\begin{figure}[H]
\centering
\includegraphics[width=.90\textwidth]{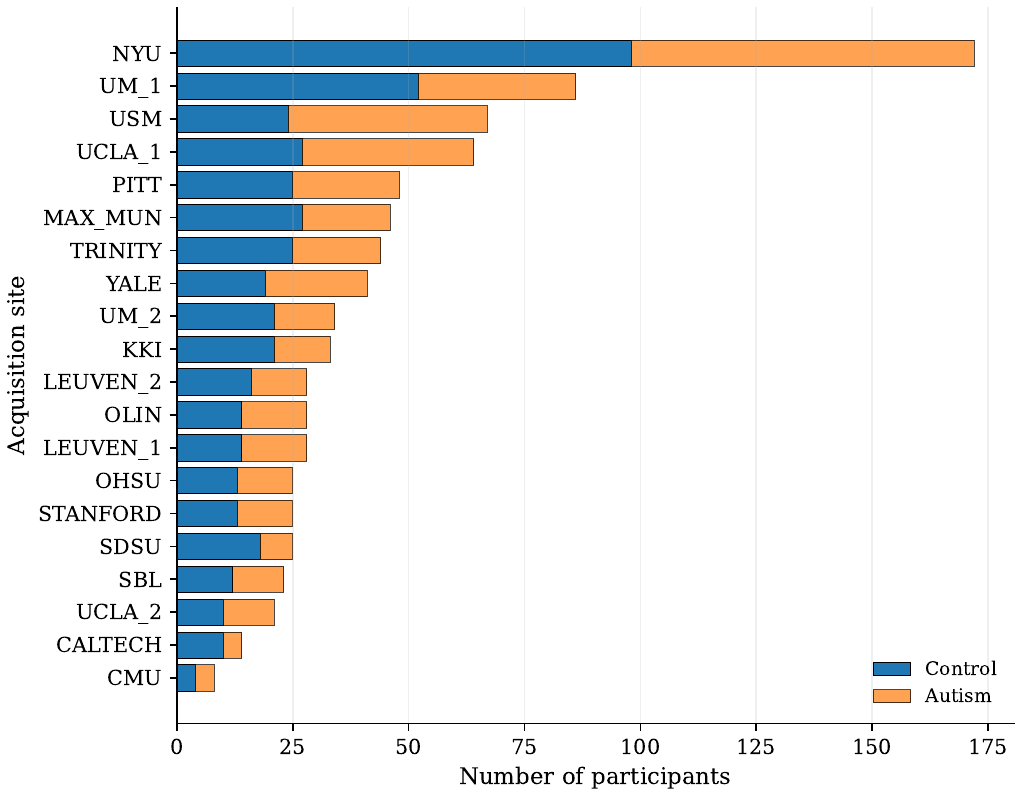}
\caption{Composition of the analytic cohort by acquisition site. Sample size and diagnostic balance differed substantially across the 20 sites.}
\label{fig:sites}
\end{figure}

The site composition was markedly unbalanced. Site totals ranged from 8 to 172 participants, and diagnostic balance also varied. This heterogeneity matters because a model trained through random subject-level splits can encounter the same site-specific acquisition characteristics during both training and testing. Such a design evaluates interpolation among known sites rather than transportability to a new center. Figure~\ref{fig:sites} displays the number of participants and diagnostic composition at each site.

For participant $i$, let $R_{ij}(t)$ denote the regional resting-state signal at time $t$ for anatomical region $j$, where $j=1,\ldots,p$ and $p=116$. Pairwise Pearson correlations produced the symmetric functional-connectivity matrix
\begin{equation}
C_i(j,k)=\Cor\{R_{ij}(t),R_{ik}(t)\}.
\label{eq:corr}
\end{equation}
Correlations were truncated away from $\pm1$ before Fisher transformation,
\begin{equation}
Z_i(j,k)=\tfrac12\log\left\{\frac{1+C_i(j,k)}{1-C_i(j,k)}\right\}.
\label{eq:fisher}
\end{equation}
The upper-triangular elements yielded 6670 unique candidate edges per participant. These features retain detailed pairwise information but substantially exceed the number of participants.

The connectivity matrix was also represented as a weighted graph. Brain regions formed the nodes, and Fisher-transformed connections formed the weighted edges. Twenty graph summaries were computed to describe signed connectivity, node-strength distributions, spectral properties, and thresholded topology. These quantities provide lower-dimensional descriptions of integration and segregation. They are not substitutes for individual connections. Instead, they provide a complementary spatial scale that can be less sensitive to local noise or site-specific perturbations.

Age, biological sex, and mean framewise displacement formed a deliberately small participant-level representation. These variables were included because they can influence both connectivity and probability calibration. They were not interpreted causally and were not intended to replace imaging information.

\begin{figure}[H]
\centering
\includegraphics[width=.92\textwidth]{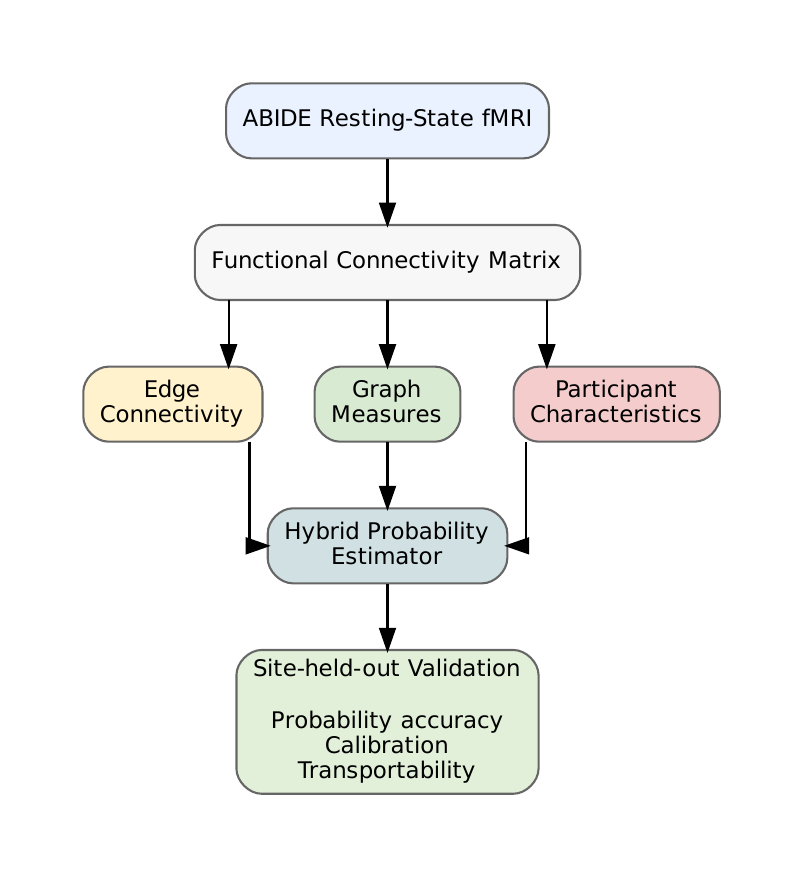}
\caption{ Overview of the proposed Hybrid Spatial Statistical Learning Framework.
Resting-state functional MRI is represented through three complementary views of brain connectivity: edge-level connectivity, graph-level network organization, and participant characteristics. Each representation contributes to a hybrid probability estimator, which is evaluated using complete site-held-out validation to assess probability accuracy, calibration, and transportability across independent imaging centers.}
\label{fig:datapipeline}
\end{figure}

The data pipeline is summarized in Figure~\ref{fig:datapipeline}. Regional time series were converted to connectivity matrices. The matrices generated a fine-scale edge representation and a coarser graph representation, while participant variables supplied an additional source of heterogeneity information. These three representations motivate the multiscale statistical framework in the next section.
The ABIDE resource provides regional time series and participant characteristics but does not provide predicted probabilities. All component probabilities and the final hybrid probability are estimated within the proposed framework using only the training sites in each cross-validation fold.

\section{Methodology}
\label{sec:methods}

The proposed framework is designed for a setting in which the most informative representation of the data may also be the least stable when prediction is transferred to a new acquisition environment. In multisite functional neuroimaging, detailed connectivity edges preserve localized and distributed signal that may be important for distinguishing autism from control participants. At the same time, a model fitted to thousands of edges can be sensitive to site-specific perturbations and may produce probabilities that are too extreme when applied to a center that was not represented during estimation. Graph summaries provide a lower-dimensional description of the same connectome. They are less detailed and may not discriminate well on their own, but they can capture broader network organization that is less sensitive to local variation. Participant characteristics provide a third source of information that helps account for variation not represented fully by the imaging data.

The Hybrid Spatial Statistical Learning Framework treats these representations as coordinated views of one outcome rather than as competing prediction problems. The edge representation is used to retain fine-scale information. The graph representation provides a broader projection of network organization. The participant representation contributes limited contextual information. Their integration is aimed at preserving discrimination while improving the numerical reliability of individualized probabilities under site shift. The framework does not impose a parametric spatial covariance model in the empirical analysis; rather, it incorporates spatial structure through edge-level relationships between anatomically defined regions and graph-level summaries of the resulting brain network.

\subsection{Multiscale statistical formulation}

Let \(Y_i\in\{0,1\}\) denote autism status for participant \(i\), and let
\(\pi_i=\Pr(Y_i=1)\) denote the corresponding conditional probability. We write

\begin{equation}
Y_i\mid\pi_i\sim\mathrm{Bernoulli}(\pi_i),
\qquad
\mathrm{logit}(\pi_i)=\eta_i,
\label{eq:latent}
\end{equation}
where \(\eta_i\) is an unobserved log-odds target. The scientific premise is that this latent quantity is informed by several representations of the same functional-connectivity system. Conceptually,

\begin{equation}
\eta_i
=
\alpha
+
\eta_i^{(E)}
+
\eta_i^{(G)}
+
\eta_i^{(D)}
+
u_{s(i)},
\label{eq:latentdecomp}
\end{equation}
where \(\eta_i^{(E)}\) denotes fine-scale connectivity information, \(\eta_i^{(G)}\) denotes broader network organization, \(\eta_i^{(D)}\) denotes participant-level information, and \(u_{s(i)}\) represents residual variation associated with acquisition site \(s(i)\).

Equation~\eqref{eq:latentdecomp} is a conceptual model for the structure of the information rather than the directly fitted estimator. Estimating a full latent spatial process for all pairwise connections would be impractical in this application because the number of connectivity features greatly exceeds the number of participants and because only 20 independent sites are available. We therefore approximate the latent formulation through separate probability estimators constructed from the three observed representations. These estimators are then integrated on the probability scale. The operational model is thus a finite-dimensional approximation motivated by Equation~\eqref{eq:latentdecomp}, not an exact fit of that additive latent logit model.

This formulation follows naturally from multiscale ideas in spatial statistics. Nonstationary covariance models show that one global dependence structure may fail to represent local variation adequately \citep{Fuentes2001,Fuentes2002Biometrika}. Spectral and multivariate methods provide related decompositions in which different components capture variation at distinct spatial resolutions \citep{Fuentes2005JASA,FuentesReichLee2008,TerresFuentes2018}. In the present setting, the edge representation retains high-frequency local information because it preserves the identity of individual connections. Graph summaries provide a lower-frequency projection of the same system by aggregating information over the network. The two representations are therefore statistically related but not interchangeable.

For participant \(i\), let \(\mathbf{x}_i\) denote the vector of Fisher-transformed connectivity edges, let \(\mathbf{g}_i\) denote the graph summaries, and let \(\mathbf{d}_i\) contain age, sex, and mean framewise displacement. The corresponding component probabilities are

\begin{align}
p_i^{(E)} &= \widehat{\Pr}\{Y_i=1\mid\mathbf{x}_i\},
\label{eq:pedge}\\
p_i^{(G)} &= \widehat{\Pr}\{Y_i=1\mid\mathbf{g}_i\},
\label{eq:pgraph}\\
p_i^{(D)} &= \widehat{\Pr}\{Y_i=1\mid\mathbf{d}_i\}.
\label{eq:pdemo}
\end{align}
Each probability is estimated using only the training sites in the corresponding validation fold. The edge component begins with 6670 candidate connections. Within the training data, edges are ranked by their univariate association with the outcome, and the 300 highest-ranking features are retained. Standardization and principal-component reduction are then estimated from the same training observations, followed by class-weighted ridge logistic regression. This component is intended to preserve detailed signal while controlling instability caused by the high-dimensional feature space.

The graph component uses the 20 summaries described in Section~\ref{sec:data}. Missing values are imputed from the training data, and the variables are standardized before fitting class-weighted ridge logistic regression. The purpose of this component is not to reproduce the edge signal. It provides a compact description of network organization that may remain more stable when local connectivity patterns change across sites.

The participant component is deliberately small. Age, sex, and mean framewise displacement are imputed and standardized within the training data before ridge logistic regression is fitted. These variables provide context for variation that may affect both connectivity and probability calibration, but they are not interpreted causally and are not intended to replace the imaging information.

The hybrid estimator is
\begin{equation}
\widehat{\pi}_i^{(H)}
=
w_Ep_i^{(E)}
+
w_Gp_i^{(G)}
+
w_Dp_i^{(D)},
\qquad
w_E+w_G+w_D=1,
\quad
w_E,w_G,w_D\geq0.
\label{eq:hybrid}
\end{equation}
Because the weights are nonnegative and sum to one, Equation~\eqref{eq:hybrid} remains a valid probability. The estimator should be interpreted as a structured convex combination of multiscale probability estimators. It is not presented as the exact likelihood estimator of Equation~\eqref{eq:latentdecomp}. Its purpose is to preserve the ordering provided by the detailed representation while allowing broader and potentially more stable representations to influence the scale of the resulting probability.

This distinction is central to the framework. A conventional ensemble may combine unrelated algorithms because their prediction errors differ. The proposed estimator instead combines models tied to prespecified representations of the same connectome. The edge, graph, and participant components are estimated under the same validation design and target the same conditional probability. Their differences arise from the scale of the information they retain, not from arbitrary algorithmic diversity.

Conditional on the fitted component models, the squared-error risk of the hybrid probability can be written as

\[
\mathbb{E}
\left[
\left\{
Y_i-\widehat{\pi}_i^{(H)}
\right\}^{2}
\mid\mathcal{T}
\right]
=
\pi_i(1-\pi_i)
+
\left[
\pi_i-
\sum_{k\in\{E,G,D\}}w_kp_i^{(k)}
\right]^2,
\]
where \(\mathcal{T}\) denotes the training data. The first term is irreducible Bernoulli variation. The second is the squared error of the estimated probability. A component can therefore improve the hybrid even if it does not improve ranking. It is useful when it moves the combined estimate closer to the true conditional probability.

This provides the statistical explanation for the role of the graph component. High-dimensional edge models can have low bias for detailed signal but high variance under site shift. Graph summaries compress the connectome and may therefore lose discrimination, yet their lower-dimensional structure can produce less variable probability estimates. A small graph contribution can reduce probability error when the edge model becomes overconfident, even if the graph model has weak standalone AUC. The proposed framework is intended for precisely this bias--variance tradeoff.

The weights are specified before evaluation on the held-out sites. In the simulation, the values are \(0.60\), \(0.25\), and \(0.15\) for the edge, graph, and participant components. In the ABIDE application, the corresponding values are \(0.80\), \(0.10\), and \(0.10\). The larger edge weight reflects the expectation that detailed connectivity carries most of the discrimination. The smaller graph and participant weights allow those representations to moderate probabilities without dominating the fine-scale signal. The selected weights reflect the relative roles of the three representations within the proposed framework rather than an attempt to maximize predictive performance. The edge-level model consistently provided the greatest discriminatory power in both preliminary analyses and the simulation study and therefore serves as the primary contributor to the hybrid probability. Graph-level summaries capture broader aspects of network organization that are less discriminative on their own but provide complementary information that improves the stability of probability estimates across heterogeneous imaging sites. Participant characteristics account for systematic demographic and acquisition-related variability that is not represented by the imaging features. The weights were therefore chosen to preserve the dominant contribution of the edge representation while allowing the complementary representations to moderate probability estimation. Because the weights were fixed before evaluation, the reported performance reflects the proposed methodology itself rather than an additional optimization over the validation data.

A learned second-stage combination is possible, but it would require additional cross-fitting. If the same outcomes were used both to estimate the component models and to learn their weights, the resulting evaluation would reuse information. Avoiding this would require a nested site-level procedure. With only 20 sites, such a design would leave limited independent information for both weight estimation and transportability assessment. The prespecified convex rule therefore favors a transparent evaluation of multiscale integration over optimization of the combination.

\subsection{Transportability and probability assessment}

Acquisition site is part of the statistical problem rather than only a nuisance variable. Participants scanned at the same center share hardware, acquisition protocols, and recruitment practices. Conceptually, the residual site contribution in Equation~\eqref{eq:latentdecomp} may be represented as

\begin{equation}
u_s\sim N(0,\sigma_u^2),
\qquad s=1,\ldots,S.
\label{eq:site}
\end{equation}
The empirical analysis does not estimate \(u_s\) for a new site because that quantity would be unavailable at the time of prediction. Instead, transportability is assessed directly by withholding complete acquisition sites during model fitting.

The 20 sites are divided into five grouped folds. In each fold, every operation that can learn from the data is estimated using the training sites only. This includes feature ranking, imputation, standardization, principal-component estimation, component-model fitting, and construction of the hybrid probability. Participants in the held-out fold are therefore evaluated using models that have not seen either those individuals or any participant from the same site group.

This design answers a different question from random subject-level cross-validation. Random splitting evaluates interpolation among sites already represented in the training sample. Complete site holdout evaluates whether the learned relationship transports to a new acquisition environment. The latter is the relevant target for multisite neuroimaging and is central to the proposed framework.

The primary inferential target is the quality of the individualized probability. We evaluate this using the Brier score,

\begin{equation}
\mathrm{BS}
=
\frac{1}{n}
\sum_{i=1}^{n}
\left(
Y_i-\widehat{\pi}_i
\right)^2,
\label{eq:brier}
\end{equation}
together with log loss and calibration curves. These measures assess the numerical agreement between estimated probabilities and observed outcomes. Discrimination is evaluated separately through ROC AUC, average precision, sensitivity, specificity, and balanced accuracy. The distinction matters because a model may preserve participant ranking while changing the scale of its probabilities substantially.

Uncertainty in the main empirical comparisons is quantified using paired participant-level bootstrap resampling of the pooled out-of-fold predictions. The same resampled participants are used for every model so that differences are evaluated on a common sample. These intervals describe uncertainty conditional on the 20 observed sites. Because the broader inferential target concerns performance across possible future sites, site-level sensitivity analyses are also considered to show how conclusions change when the acquisition environment is treated as the unit of variation.

The final framework is therefore unified by one statistical target and one validation principle. The three component models estimate the same individualized autism probability from different spatial representations. Their convex integration is motivated by the complementary bias and variance properties of those representations. Their evaluation is carried out entirely at unseen sites. Section~\ref{sec:simulation} examines when this multiscale integration should improve probability accuracy, and Section~\ref{sec:application} evaluates whether the same mechanism appears in the ABIDE data.

\section{Simulation study}
\label{sec:simulation}

The simulation study was designed to determine what the proposed framework contributes beyond its individual components. The central question was whether a model built from detailed local features can retain its ability to distinguish outcomes while producing more reliable probabilities when it is combined with broader and more stable representations. We studied this question under increasing heterogeneity across sites because transportability to a new acquisition environment is the principal challenge addressed by the framework.

\begin{figure}[H]
\centering
\includegraphics[width=.82\textwidth]{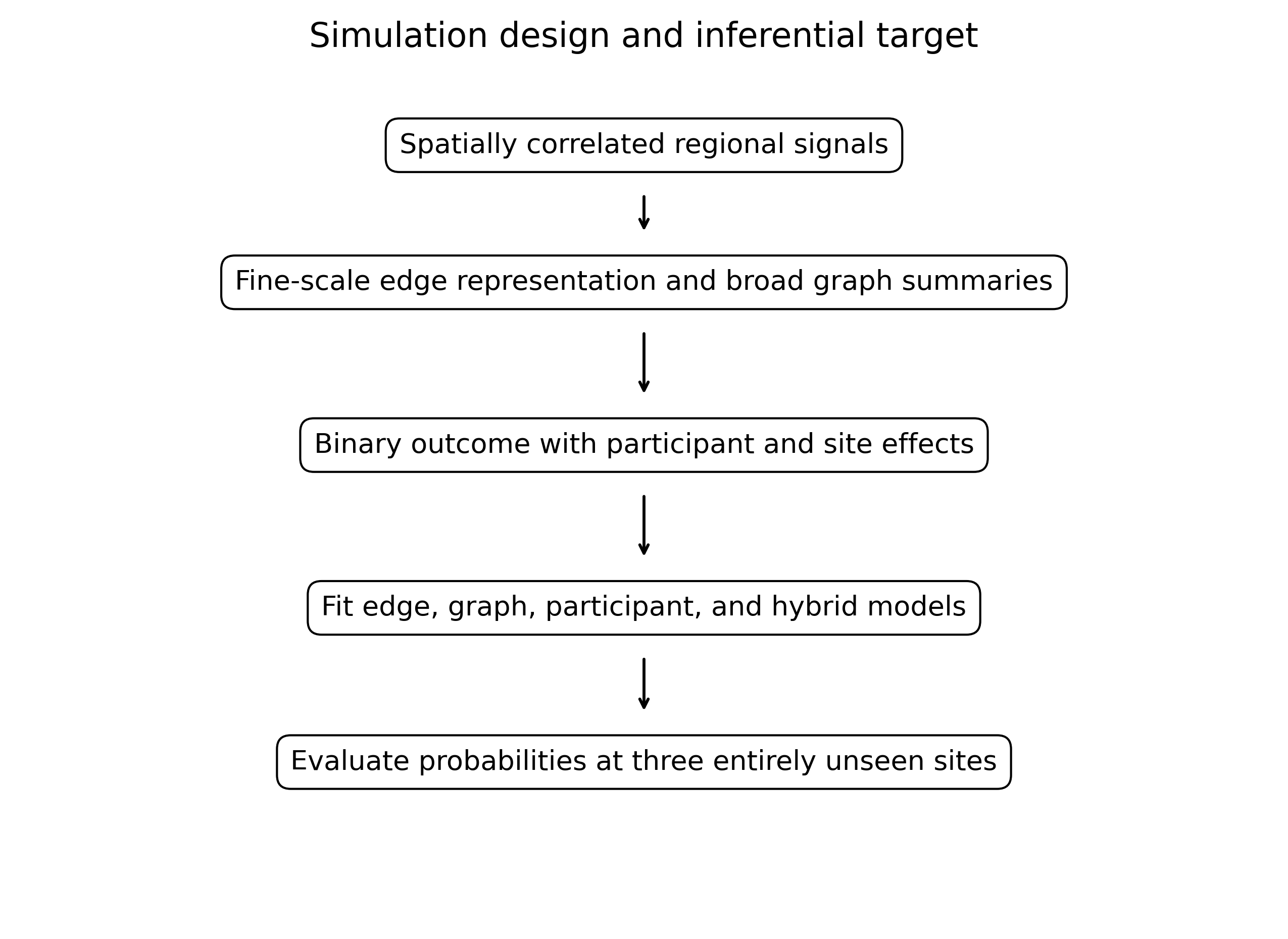}
\caption{Simulation design. Spatially correlated regional measurements generate fine-scale edge variables and broader graph summaries. Outcomes depend on information from both scales and on a site-specific effect. Models are fitted without data from three sites and are evaluated on those unseen acquisition environments.}
\label{fig:simdesign}
\end{figure}

The study therefore had three related aims. We first examined whether the edge model remained the strongest source of discrimination when the outcome depended primarily on fine-scale information. We then assessed whether graph and participant information could improve the numerical accuracy of the estimated probabilities even when those components were weaker classifiers on their own. Finally, we evaluated whether any advantage of the hybrid became more apparent as the sites differed more strongly from one another. Figure~\ref{fig:simdesign} summarizes the design and the inferential target.

\subsection{Models compared}

The edge model represents the detailed local view of the simulated system. It is fitted to the full set of edge-like predictors after screening and regularization. This model is expected to provide the strongest discrimination because the data-generating mechanism places its largest effects on two local variables. It also faces the greatest estimation challenge because it is fitted in a comparatively high-dimensional feature space.

The graph model represents the broader view. It uses a small set of summaries derived from the same regional process and edge representation. These summaries compress local information into measures of overall level and variability. The graph model is not expected to reproduce the edge model's discrimination. Its role is to capture lower-dimensional structure that may remain more stable when the distribution changes across sites.

A participant model uses one continuous covariate and one binary covariate. It provides a limited source of nonspatial information and mirrors the role played by age, sex, and motion in the ABIDE analysis. The hybrid model combines the probabilities from these three components using weights of \(0.60\), \(0.25\), and \(0.15\). The weights were fixed before evaluating the simulations. The edge component receives the largest weight because fine-scale signal drives the outcome. The graph and participant components retain meaningful influence without being allowed to dominate the probability estimate.

The comparison is therefore not between four unrelated algorithms. It is between three representations of one simulated process and a hybrid that integrates them. The simulation asks whether the broader components can improve probability quality even when the edge component remains the best standalone classifier.

\subsection{Simulation design}

We generated \(R=18\) regional measurements for each participant at equally spaced locations on a unit circle. The circle provides a simple domain with a transparent distance structure and avoids placing some regions at artificial boundaries. It is not intended to reproduce cortical anatomy. Its purpose is to create correlated regional measurements from which local and broad representations can be derived.

For participant \(i\), the regional vector was generated as
\begin{equation}
\mathbf{Z}_i\sim N_R(\mathbf{0},\boldsymbol{\Sigma}),
\label{eq:simgp}
\end{equation}
with
\begin{equation}
\Sigma_{jk}
=
\exp\!\left\{-\frac{d(s_j,s_k)}{0.70}\right\}
+
0.10I(j=k),
\label{eq:simcov}
\end{equation}
where \(d(s_j,s_k)\) is Euclidean distance between locations. The exponential form creates a clear decay in dependence as regions become farther apart. The value \(0.70\) was chosen relative to the unit-circle geometry. Adjacent regions are separated by about \(0.35\) units, which gives a spatial covariance of approximately \(0.61\), whereas regions at opposite sides of the circle are separated by \(2\) units, giving a covariance of approximately \(0.06\). The process therefore has meaningful local dependence without making all regions nearly collinear.

The added variance of \(0.10\) represents modest region-specific variation that is not shared with neighboring locations. Because the spatial process variance is one, this term contributes about \(9\%\) of the total marginal variance. It prevents the simulated field from being unrealistically smooth. Neither value was estimated from ABIDE. They were selected to create an interpretable setting in which local dependence is present and the broad summaries remain informative.

The covariance matrix was factored as
\[
\boldsymbol{\Sigma}=\mathbf{L}\mathbf{L}^{T},
\]
and the regional vector was generated from
\[
\mathbf{Z}_i=\mathbf{L}\boldsymbol{\varepsilon}_i,
\qquad
\boldsymbol{\varepsilon}_i\sim N_R(\mathbf{0},\mathbf{I}).
\]
Independent participant vectors generated in this way share the covariance in Equation~\eqref{eq:simcov}.

Fine-scale edge variables were constructed from noisy pairwise products,
\begin{equation}
X_{ijk}
=
Z_i(s_j)Z_i(s_k)+e_{ijk},
\qquad j<k,
\label{eq:simedges}
\end{equation}
where \(e_{ijk}\sim N(0,0.80^2)\). The pairwise products create local variables whose values depend jointly on two regions. They are not intended to be literal sample correlations. They provide a controlled analogue of connectivity edges in which many predictors are generated from a common spatial process.

Four graph summaries were calculated from the same simulated data. They described the mean and standard deviation of the regional vector and the mean and upper-tail magnitude of the absolute edge values. These variables compress the local measurements into a small broad-scale representation. The edge and graph features therefore contain related information, but at different resolutions.

Each data set contained 12 sites with 30 participants per site. Sites 10 through 12 were reserved for evaluation, leaving nine sites for model fitting. The outcome was generated from
\begin{equation}
\begin{aligned}
Y_i\mid\pi_i &\sim \mathrm{Bernoulli}(\pi_i),\\
\mathrm{logit}(\pi_i)
&=0.65X_{i,4}-0.55X_{i,21}+0.45G_{i,1}\\
&\quad-0.35G_{i,3}+0.20D_{i,1}+u_{s(i)}.
\end{aligned}
\label{eq:simoutcome}
\end{equation}
The linear predictor was centered within each simulated data set before outcomes were drawn. The two edge effects are largest, so the edge model is expected to carry most of the discrimination. The graph and participant effects are smaller but contribute to the true conditional probability. This construction directly reflects the mechanism that the hybrid is intended to exploit.

The site effects were generated independently as
\[
u_s\sim N(0,\sigma_u^2).
\]
We used \(\sigma_u=0.20\), \(0.60\), and \(1.00\) to represent low, moderate, and high heterogeneity. Only this variance changed across scenarios. The first setting produces relatively small site shifts. The second creates clearly visible differences in baseline propensity. The third makes the site effect large relative to the participant-level signal. Each scenario was repeated 40 times.

The component models were fitted on the nine training sites. The edge model used univariate screening to retain 25 variables, followed by standardization and ridge logistic regression. The graph and participant models used standardized predictors and ridge logistic regression. All preprocessing was estimated from the training sites only. The hybrid probability was formed from the fixed weights described above and evaluated on the three unseen sites.

The estimand was the conditional probability of the binary outcome at those unseen sites. ROC AUC measured discrimination. The Brier score and log loss measured probability accuracy. A lower Brier score or log loss indicates that the predicted probabilities are closer to the outcomes generated by Equation~\eqref{eq:simoutcome}. Means and Monte Carlo standard deviations were calculated across the 40 replications.

\subsection{Results}

\begin{figure}[H]
\centering
\includegraphics[width=.72\textwidth]{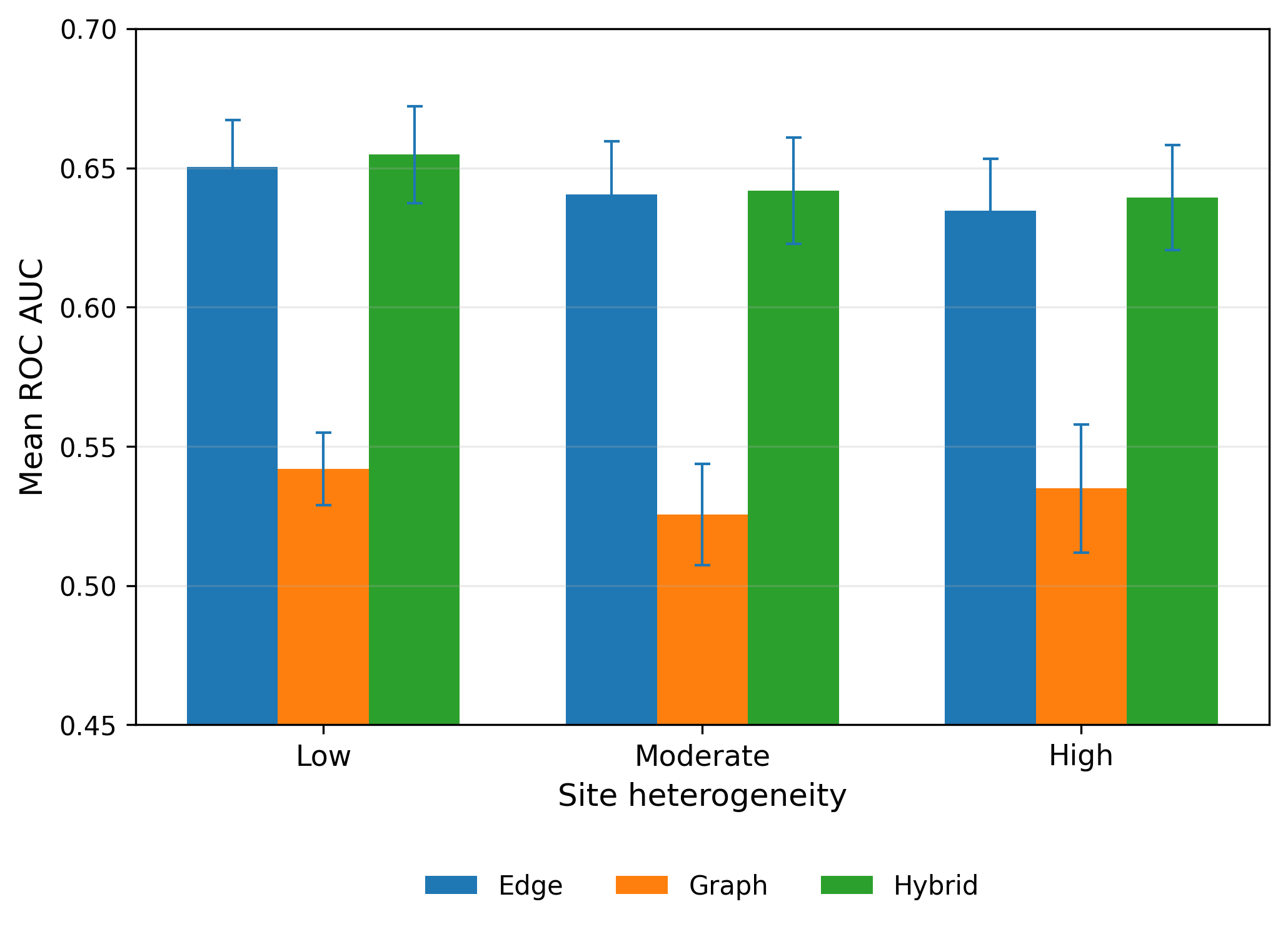}
\caption{Discrimination at three entirely unseen sites. Bars show mean ROC AUC across 40 replications, and error bars show normal-approximation 95\% intervals for the mean. The hybrid retains the discrimination of the edge model as site heterogeneity increases.}
\label{fig:simauc}
\end{figure}

The first result concerns discrimination. As intended, the edge model was the strongest standalone component. Its mean ROC AUC ranged from \(0.635\) under high site heterogeneity to \(0.650\) under low heterogeneity. The graph model was much weaker, with mean AUC between \(0.526\) and \(0.542\). The hybrid retained the ordering supplied by the edge model. Its mean AUC ranged from \(0.639\) to \(0.655\), and the differences from the edge model were small in all three settings. Figure~\ref{fig:simauc} shows that combining the components did not sacrifice the fine-scale discrimination.

\begin{figure}[H]
\centering
\includegraphics[width=.72\textwidth]{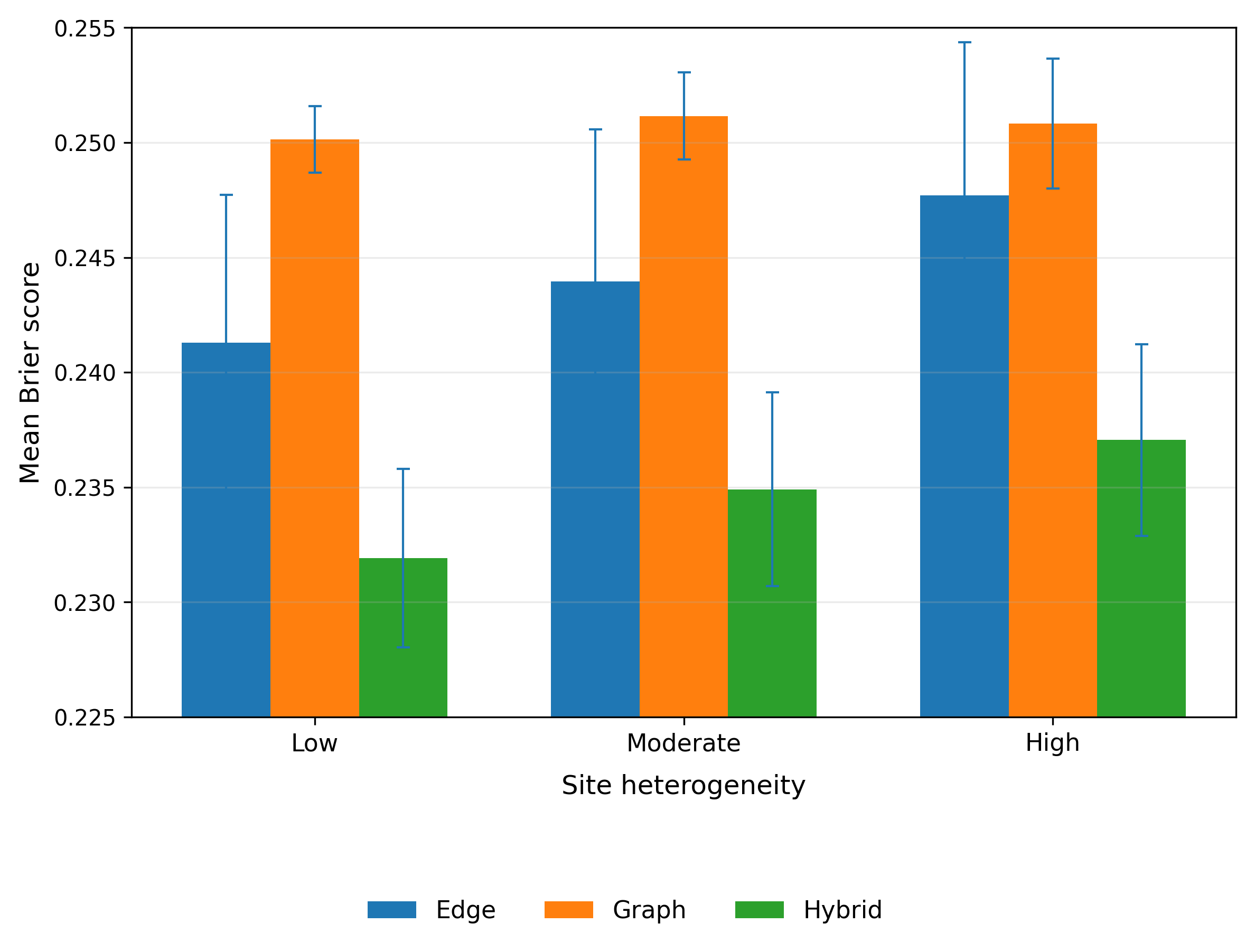}
\caption{Probability accuracy at three entirely unseen sites. Bars show mean Brier score across 40 replications, and error bars show normal-approximation 95\% intervals for the mean. Lower values are better. The hybrid improves probability accuracy while preserving edge-model discrimination.}
\label{fig:simbrier}
\end{figure}

The second result concerns probability accuracy. The hybrid had the lowest mean Brier score at every level of heterogeneity. Under low heterogeneity, the score decreased from \(0.241\) for the edge model to \(0.232\) for the hybrid. Under moderate heterogeneity, it decreased from \(0.244\) to \(0.235\). Under high heterogeneity, it decreased from \(0.248\) to \(0.237\). Log loss showed the same pattern. Figure~\ref{fig:simbrier} displays the Brier-score improvement across the three heterogeneity settings, and Table~\ref{tab:simulation} provides the corresponding ROC AUC, Brier score, and log-loss results with Monte Carlo standard deviations. The advantage of the hybrid therefore did not arise from a stronger ranking of participants. It arose because the combined probabilities were numerically closer to the observed outcomes.

\begin{table}[H]
\centering
\small
\caption{Simulation performance at three completely held-out sites. Values are means across 40 replications, with Monte Carlo standard deviations in parentheses.}
\label{tab:simulation}
\begin{tabular}{llccc}
\toprule
Site heterogeneity & Model & ROC AUC & Brier score & Log loss\\
\midrule
Low
& Edge   & 0.650 (0.055) & 0.241 (0.021) & 0.686 (0.052)\\
& Graph  & 0.542 (0.042) & 0.250 (0.005) & 0.694 (0.010)\\
& Hybrid & 0.655 (0.056) & 0.232 (0.013) & 0.656 (0.027)\\
\addlinespace
Moderate
& Edge   & 0.640 (0.062) & 0.244 (0.021) & 0.690 (0.051)\\
& Graph  & 0.526 (0.059) & 0.251 (0.006) & 0.696 (0.013)\\
& Hybrid & 0.642 (0.062) & 0.235 (0.014) & 0.662 (0.029)\\
\addlinespace
High
& Edge   & 0.635 (0.061) & 0.248 (0.021) & 0.702 (0.053)\\
& Graph  & 0.535 (0.074) & 0.251 (0.009) & 0.695 (0.019)\\
& Hybrid & 0.639 (0.061) & 0.237 (0.013) & 0.667 (0.028)\\
\bottomrule
\end{tabular}
\end{table}

\begin{figure}[H]
\centering
\includegraphics[width=.72\textwidth]{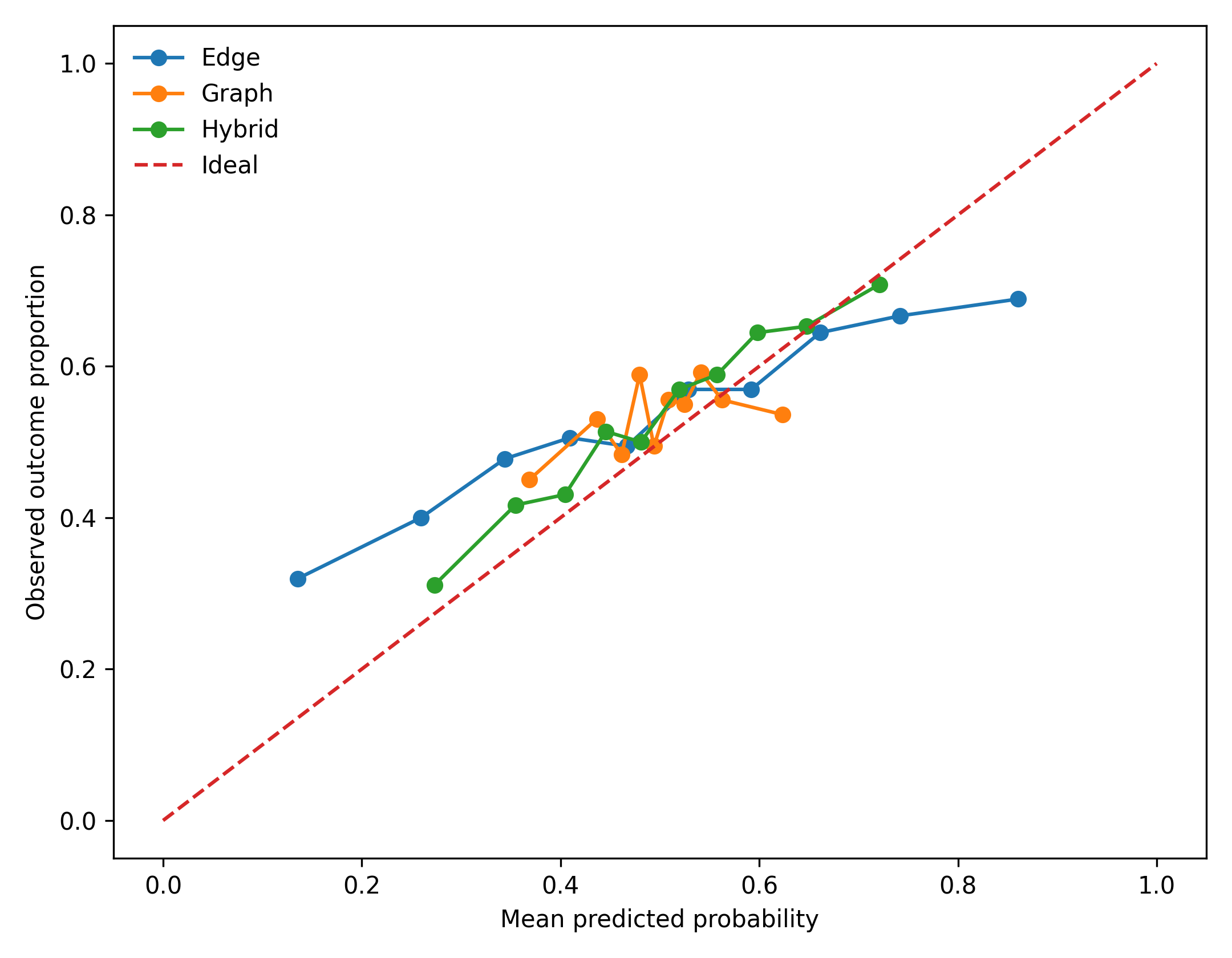}
\caption{Calibration under high site heterogeneity, pooled across the 40 replications. The diagonal denotes ideal calibration. The hybrid moderates the edge probabilities while preserving substantially more discrimination than the graph model.}
\label{fig:simcalibration}
\end{figure}

The calibration curves under high heterogeneity provide a direct view of this distinction. As shown in Figure~\ref{fig:simcalibration}, the edge model produces a wider range of probabilities and is more prone to confident errors when the held-out sites differ from the training sites. The graph model stays closer to the center but lacks discrimination. The hybrid lies between these behaviors. It preserves much of the ordering supplied by the edge model while moving the probability scale closer to the observed outcome frequencies and closer to the ideal calibration line.

These findings clarify what is learned from the simulation. The graph component is not valuable because it becomes a better classifier than the edge component. It is valuable because it supplies a broader and less variable view of the same process. When the acquisition environment changes, this information can reduce the probability error of the detailed model. The hybrid therefore improves the scale of the prediction more than its rank.

The simulation also shows why increasing site heterogeneity matters. As \(\sigma_u\) increases, the edge model continues to identify much of the participant ordering, but its Brier score and log loss deteriorate. The hybrid is less sensitive to that change. This is precisely the setting for which the framework was developed: strong local signal is present, but transport to a new site affects the reliability of the probability estimate.

The study is intentionally focused. It does not reproduce the full temporal dynamics of rs-fMRI or claim that pairwise products are equivalent to estimated functional correlations. Its purpose is to isolate the statistical mechanism of the method. Under a known spatial dependence structure and a controlled site shift, the hybrid preserves fine-scale discrimination and improves out-of-site probability accuracy. The ABIDE application in Section~\ref{sec:application} examines whether the same pattern appears in a substantive multisite neuroimaging study.

\section{Application to multisite autism neuroimaging}
\label{sec:application}

The empirical analysis was designed to determine whether the proposed multiscale framework improves individualized autism probability estimation when the model is transported across imaging centers. The principal comparison is therefore not whether the hybrid produces a visibly different ROC curve. The hybrid is intended to preserve the participant ordering supplied by detailed connectivity while reducing the probability error that arises when an edge model is applied to a new acquisition environment.

Relative to the edge model, the hybrid reduced the Brier score from \( 0.256 \) to \( 0.243 \), an improvement of \( 0.012 \), or \( 4.9\% \). Log loss decreased from \( 0.733 \) to \( 0.684 \), an improvement of \( 0.049 \), or \( 6.6\% \). ROC AUC changed only from \( 0.646 \) to \( 0.647 \). Thus, the principal empirical gain is improved probability quality while discrimination is preserved.

\subsection{Application of the framework}

The edge, graph, and participant representations were constructed independently within each training fold. The edge model used Fisher-transformed functional connections after training-fold feature ranking, standardization, and principal-component reduction. Class-weighted ridge logistic regression produced the edge probability. This representation retained detailed information about specific regional relationships and supplied most of the discrimination.

The graph model used the 20 network summaries described in Section~\ref{sec:data}. Missing values were imputed from the training sample, and the variables were standardized before fitting class-weighted ridge logistic regression. The participant model used age, sex, and mean framewise displacement as contextual information.

The hybrid probability was formed using Equation~\eqref{eq:hybrid} with weights \(0.80\), \(0.10\), and \(0.10\) for the edge, graph, and participant components. Five grouped folds were used, with complete acquisition sites held out in each fold. Every learned operation was repeated using the training sites only.

\subsection{The hybrid preserves discrimination and improves probability accuracy}

Table~\ref{tab:application_performance} reports the site-held-out results using one common participant-level prediction file as the source of truth.

\begin{table}[H]
\centering
\small
\caption{Site-held-out model performance. Lower Brier score and log loss indicate better probability estimates.}
\label{tab:application_performance}
\begin{tabular}{lccc}
\toprule
Metric & Edge & Graph & Hybrid\\
\midrule
Balanced accuracy & 0.605 & 0.466 & 0.608\\
Sensitivity & 0.665 & 0.524 & 0.668\\
Specificity & 0.544 & 0.408 & 0.549\\
ROC AUC & 0.646 & 0.469 & 0.647\\
Average precision & 0.585 & 0.442 & 0.587\\
Brier score & 0.256 & 0.255 & 0.243\\
Log loss & 0.733 & 0.704 & 0.684\\
\bottomrule
\end{tabular}
\end{table}

\begin{figure}[H]
\centering
\includegraphics[width=.74\textwidth]{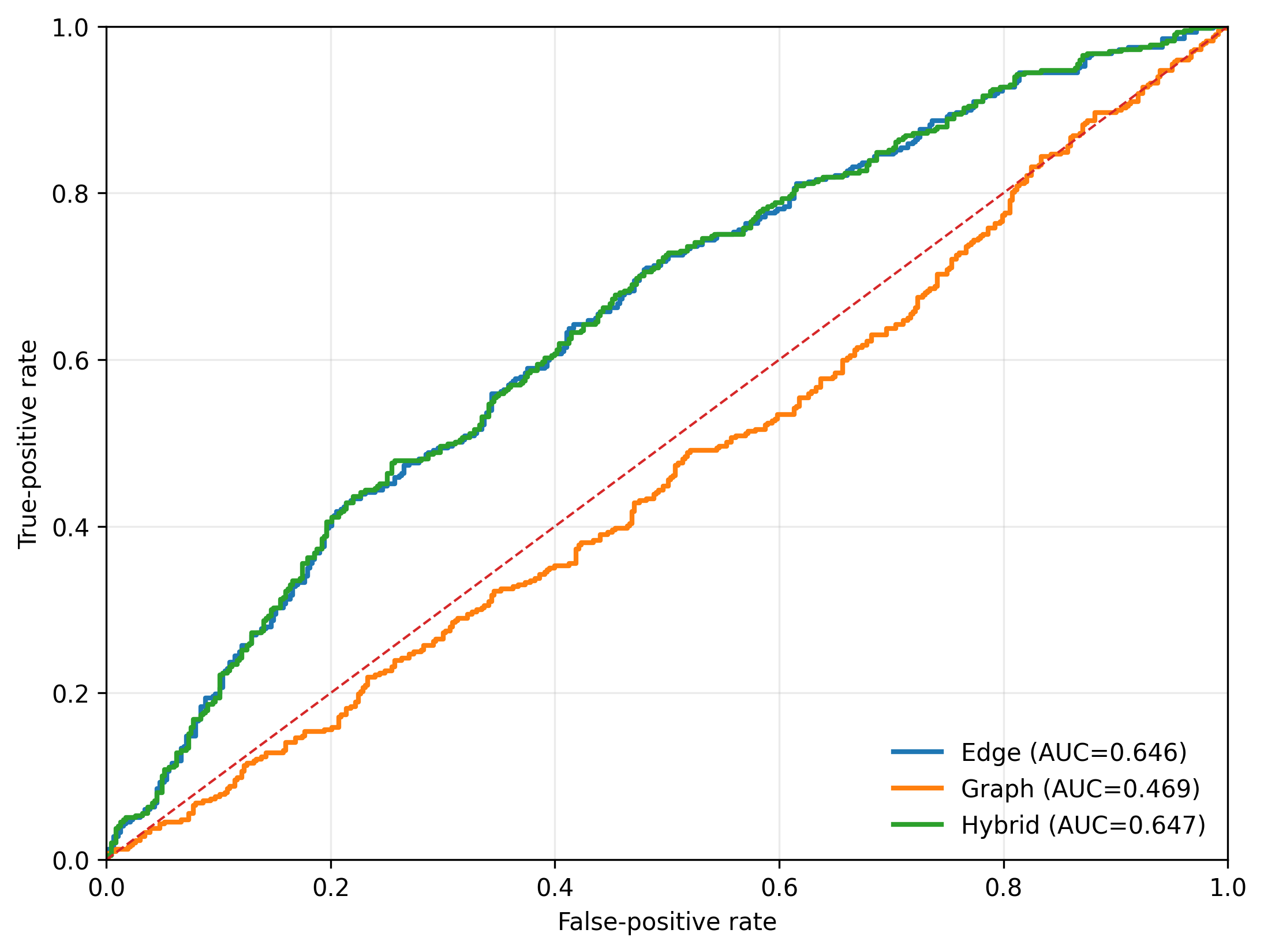}
\caption{Receiver operating characteristic curves from pooled site-held-out predictions. The edge and hybrid curves are nearly identical because the hybrid changes probability magnitude more than participant ranking.}
\label{fig:application_roc}
\end{figure}

The edge and hybrid ROC curves in Figure~\ref{fig:application_roc} are almost indistinguishable. This confirms that the hybrid retains the fine-scale ordering supplied by the edge model.

The gain is most evident on the probability scale. Table~\ref{tab:hybrid_gains} reports the absolute and relative reductions for the hybrid relative to the edge model.

\begin{table}[H]
\centering
\small
\caption{Improvement in probability accuracy for the hybrid relative to the edge model. Positive reductions indicate better performance for the hybrid.}
\label{tab:hybrid_gains}
\begin{tabular}{lccc}
\toprule
Metric & Absolute reduction & Relative reduction & 95\% interval\\
\midrule
Brier score & 0.012 & 4.9\% & 0.009 to 0.016\\
Log loss & 0.049 & 6.6\% & 0.035 to 0.064\\
\bottomrule
\end{tabular}
\end{table}

\begin{figure}[H]
\centering
\includegraphics[width=.74\textwidth]{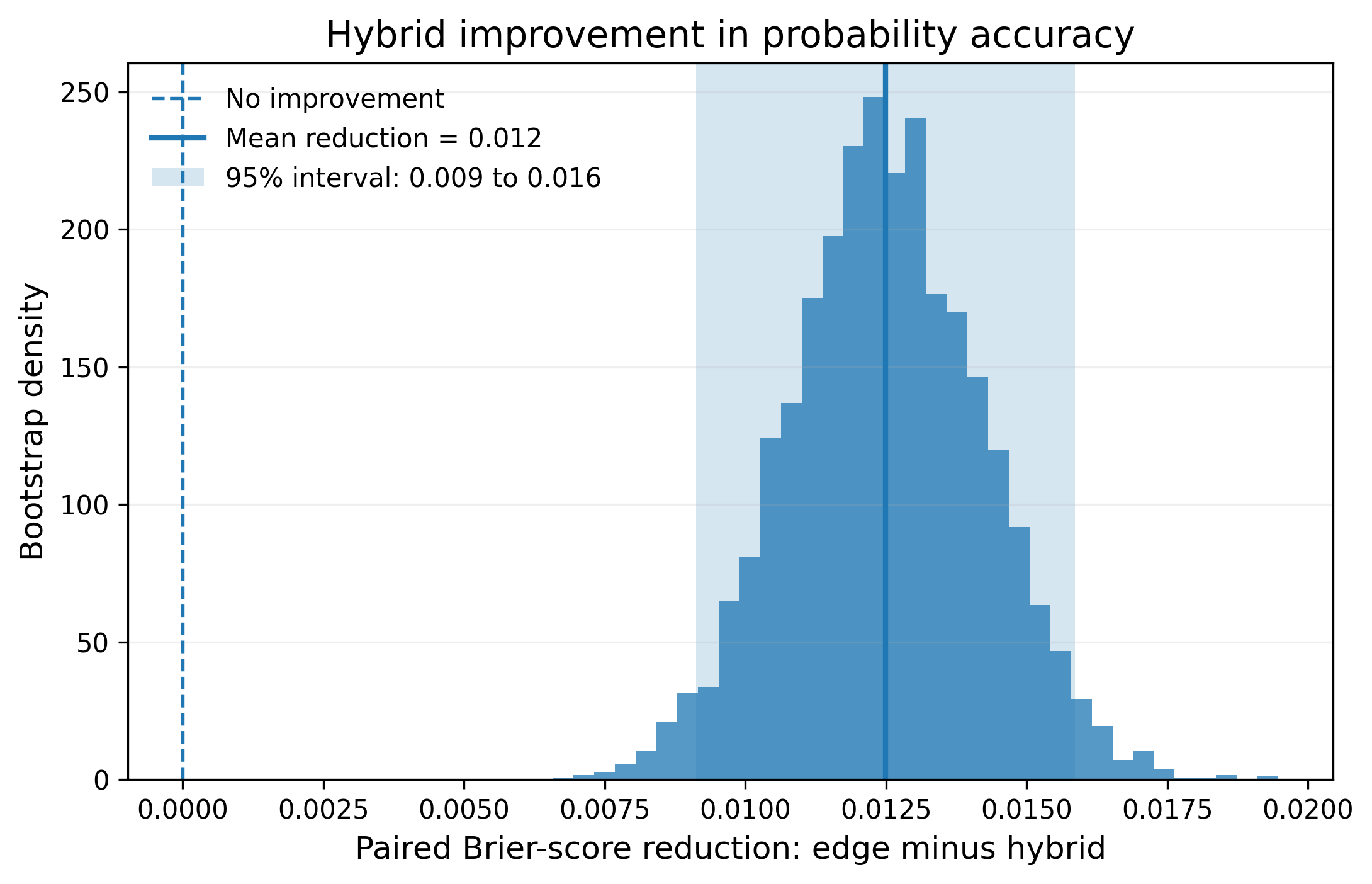}
\caption{Participant-bootstrap distribution of the paired Brier-score reduction for the hybrid relative to the edge model. Positive values indicate lower probability error for the hybrid. The shaded region is the 95\% bootstrap interval, and the dashed line marks no improvement.}
\label{fig:hybrid_brier}
\end{figure}

\begin{figure}[H]
\centering
\includegraphics[width=.74\textwidth]{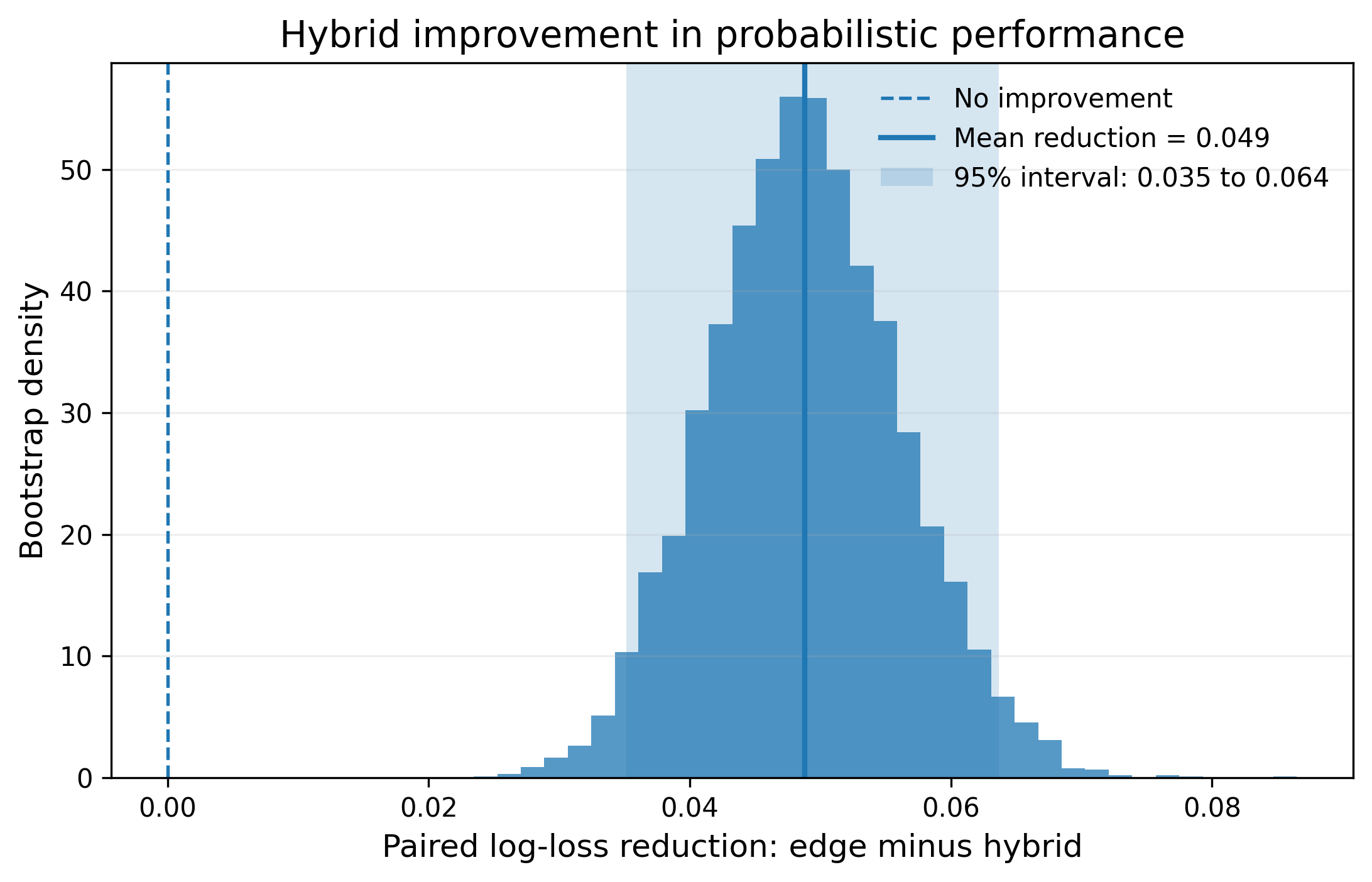}
\caption{Participant-bootstrap distribution of the paired log-loss reduction for the hybrid relative to the edge model. Positive values indicate better probabilistic performance for the hybrid. The shaded region is the 95\% bootstrap interval, and the dashed line marks no improvement.}
\label{fig:hybrid_logloss}
\end{figure}

Figures~\ref{fig:hybrid_brier} and \ref{fig:hybrid_logloss} display the paired bootstrap distribution of the improvement for the hybrid relative to the edge model. Positive values indicate that the hybrid has lower probability error. Because the comparison is paired at the participant level, these figures reveal the gain directly and avoid the misleading visual overlap that occurs when separate model intervals are plotted.

\begin{figure}[H]
\centering
\includegraphics[width=.74\textwidth]{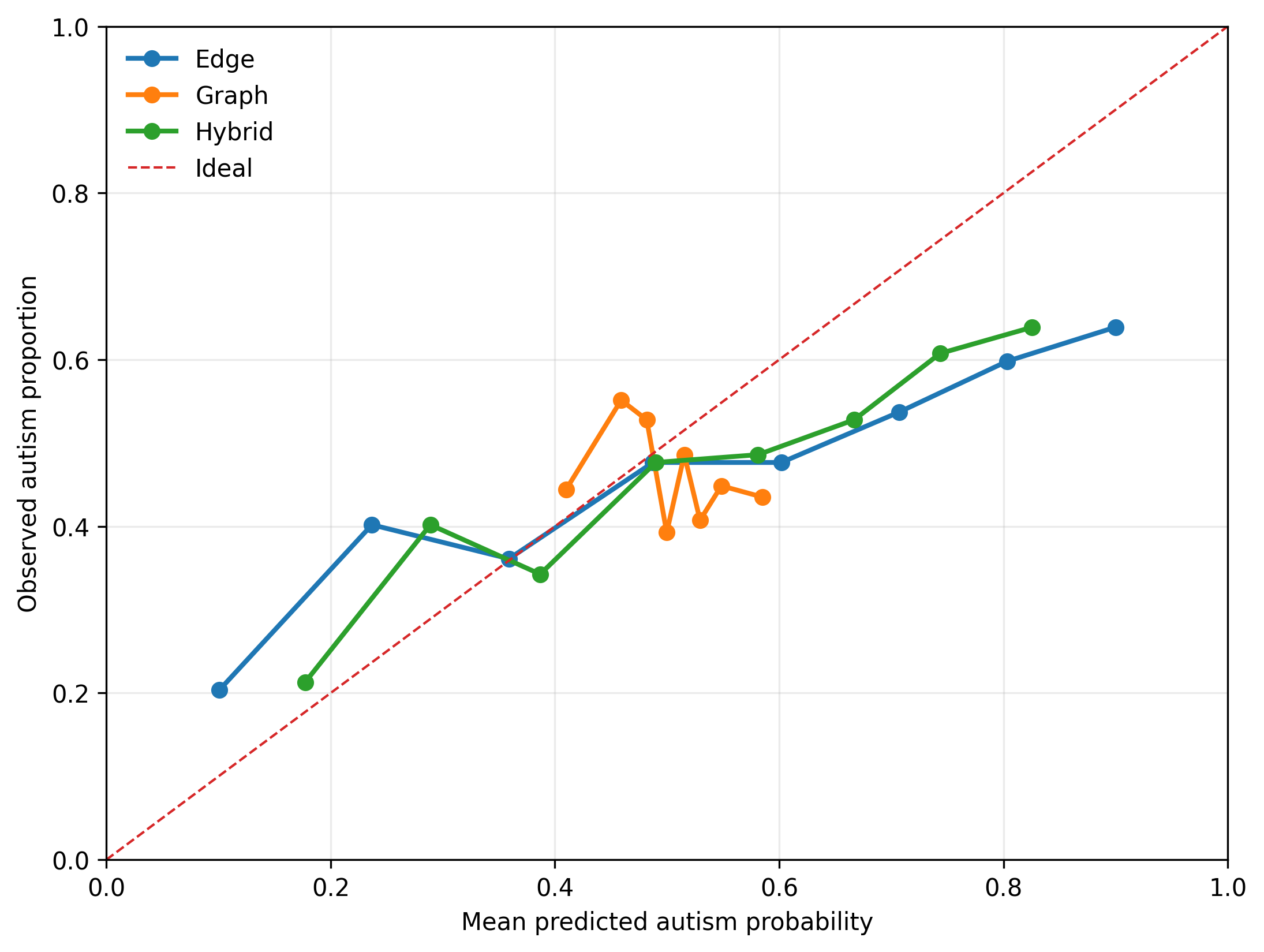}
\caption{Calibration of pooled site-held-out predictions. The diagonal represents ideal agreement between predicted autism probability and observed autism proportion.}
\label{fig:application_calibration}
\end{figure}

The calibration curves in Figure~\ref{fig:application_calibration} provide a complementary interpretation. The edge model produces a wider range of probabilities and is more vulnerable to confident errors. The graph model remains closer to the center but provides little discrimination. The hybrid preserves much of the edge ordering while moderating probability magnitude.

\subsection{The improvement across acquisition sites}

\begin{figure}[H]
\centering
\includegraphics[width=.88\textwidth]{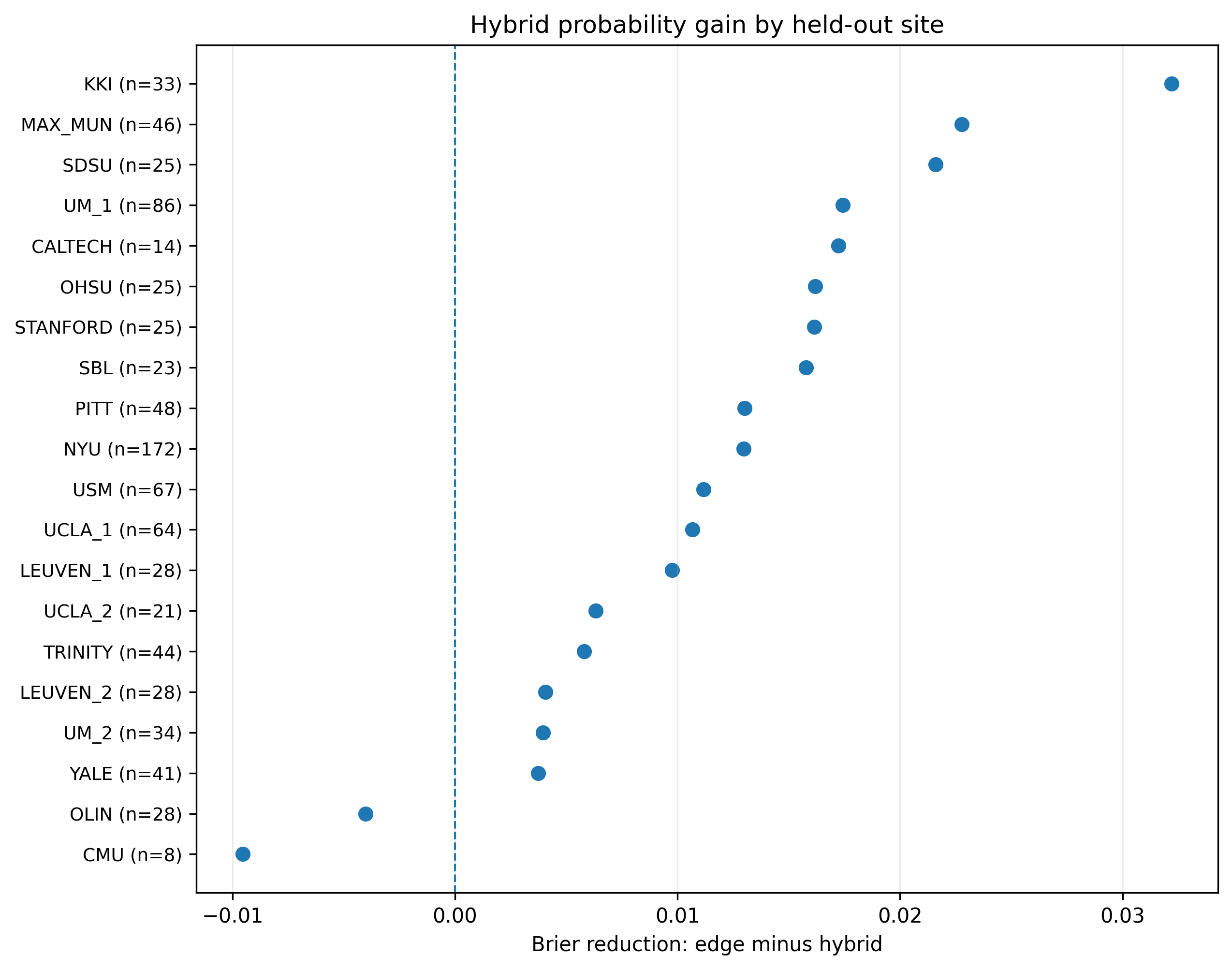}
\caption{Site-specific reduction in Brier score for the hybrid relative to the edge model. Positive values indicate that the hybrid produces more accurate probabilities. Site sample sizes are shown in parentheses.}
\label{fig:site_brier_gain}
\end{figure}

The pooled gain could be driven by only a few centers. Figure~\ref{fig:site_brier_gain} therefore shows the site-specific difference between edge and hybrid Brier scores. Positive values indicate a lower probability error for the hybrid.

\subsection{Scientific implications for autism}

The application does not identify one definitive connectivity biomarker for autism. The moderate AUC indicates that the imaging signal is distributed and heterogeneous. Detailed connections contain most of the transportable diagnostic ordering, but that ordering alone does not yield the most reliable probabilities across sites.

The hybrid result provides a more specific scientific interpretation. Fine-scale connectivity and broader network organization do not contribute in the same way. The local representation supplies discrimination. The broader representations help determine how much confidence should be placed in that signal when the acquisition environment changes. This division of roles is consistent with autism as a condition involving distributed differences in communication among brain systems rather than one isolated abnormality.

The probabilities produced here are not intended to replace clinical assessment. Their value lies in quantifying the strength of the imaging evidence and in showing how that evidence changes across participants and centers.

\subsection{Implications for multisite neuroimaging}

The main empirical conclusion is not that the hybrid creates a better classifier. It is that a multiscale probability model can retain fine-scale discrimination while reducing probability error under site shift. This distinction is relevant throughout multisite neuroimaging, where a high-dimensional representation may be discriminative but unstable and a lower-dimensional representation may be less discriminative but more robust.

The ABIDE analysis therefore supports the central premise of the paper. The strongest representation for participant ranking is not necessarily the representation that produces the most reliable probabilities. By integrating detailed connectivity with broader network organization, the Hybrid Spatial Statistical Learning Framework produces a more stable description of individualized autism probability across heterogeneous imaging environments.

\section{Discussion}
\label{sec:discussion}

This work was motivated by a statistical problem that extends well beyond autism neuroimaging. Modern biomedical studies increasingly combine high-dimensional measurements collected across multiple institutions, imaging centers, and study populations. These collaborative efforts provide unprecedented opportunities for understanding complex diseases, but they also introduce substantial heterogeneity arising from differences in acquisition protocols, instrumentation, preprocessing pipelines, and participant characteristics. Developing statistical methods that remain reliable under these conditions has become an important challenge in biostatistics, particularly when the objective is individualized probability estimation rather than classification alone.

The Hybrid Spatial Statistical Learning Framework proposed in this paper addresses this challenge by recognizing that disease-related information is naturally expressed across multiple spatial scales. Rather than treating edge-based connectivity, graph-level organization, and participant characteristics as competing sources of information, the proposed methodology models them as complementary statistical representations of the same individualized probability target. Local functional connections preserve detailed regional interactions that are important for discrimination, whereas graph summaries characterize broader patterns of network organization that may be more stable across heterogeneous imaging environments. Participant-level characteristics provide additional contextual information that complements, but does not replace, the imaging data. Integrating these representations within a common probabilistic model produces individualized probabilities that remain interpretable while improving robustness to multisite heterogeneity.

This perspective differs from much of the existing neuroimaging literature. Many prediction studies have concentrated on developing increasingly sophisticated machine learning algorithms with the primary objective of maximizing classification accuracy. Penalized regression, support vector machines, random forests, graph neural networks, and deep learning approaches have substantially advanced predictive modeling in biomedical imaging. Nevertheless, relatively less attention has been devoted to the statistical quality of estimated probabilities, their calibration, and their transportability across acquisition sites. Reliable probabilities are often more informative than classifications alone because they quantify the uncertainty associated with each individual prediction.

The simulation study clarifies the statistical mechanism through which the proposed framework operates. The improvement observed for the hybrid estimator did not arise because the graph representation became a stronger classifier than the detailed connectivity model. Instead, the broader graph summaries contributed complementary information that moderated the probabilities generated by the high-dimensional edge model when data were transferred to previously unseen sites. Consequently, the principal improvements appeared in proper scoring rules such as the Brier score and log loss, while discrimination remained essentially unchanged. Different representations of the same dependent process therefore contribute differently to prediction, with some improving discrimination and others improving the reliability of individualized probabilities.

The empirical application demonstrates that the same phenomenon is present in multisite autism neuroimaging. Detailed functional connectivity contained most of the information required to distinguish autistic participants from neurotypical controls, whereas graph-level summaries primarily improved the calibration and stability of individualized probabilities. These findings support an increasingly accepted view of autism as a disorder involving distributed alterations in communication among multiple functional systems rather than abnormalities confined to isolated brain regions. They also reinforce the substantial heterogeneity that characterizes autism. The moderate discrimination observed across independent imaging centers suggests that there is no single functional connectivity signature that consistently characterizes every individual with autism. Statistical models that integrate information across multiple spatial scales therefore provide a more realistic representation of the biological complexity underlying the disorder.

The application also illustrates the importance of evaluating prediction models under realistic conditions. Validation based on random subject-level cross-validation frequently allows information from the same acquisition environment to appear in both the training and testing sets, leading to optimistic assessments of predictive performance. By validating exclusively on completely held-out imaging sites, this study evaluates a considerably more demanding problem that more closely resembles future applications of neuroimaging biomarkers. The resulting performance estimates therefore provide a more realistic assessment of model generalizability and emphasize that transportability should be considered a central component of methodological evaluation in multisite biomedical studies.

From a methodological perspective, the proposed framework contributes to the growing interface between spatial statistics and statistical learning. Functional connectivity represents only one example of a complex multiscale dependent process. Similar statistical challenges arise in spatial transcriptomics, digital pathology, molecular imaging, environmental health, ecological monitoring, and many other biomedical settings where information is distributed across multiple spatial resolutions. The methodology developed here demonstrates how complementary representations of dependent data can be combined within a coherent probabilistic framework that preserves statistical interpretability while remaining computationally practical for large-scale applications.

An additional contribution of this work is its emphasis on probability estimation as the primary inferential target. Biomedical decisions frequently depend not only on whether an individual is classified correctly, but also on the degree of confidence associated with that prediction. A model that produces well-calibrated probabilities allows investigators to distinguish between participants for whom the imaging evidence is compelling and those for whom the evidence is substantially more uncertain. This distinction is particularly important for heterogeneous disorders such as autism, where biological variability is expected rather than exceptional.

Several limitations should be acknowledged. The hybrid weights were specified a priori rather than estimated from the data in order to evaluate the statistical contribution of multiscale integration independently of weight optimization. Alternative adaptive weighting strategies represent an interesting extension. Likewise, the graph representation considered here was intentionally based on conventional network summaries. More sophisticated graph embeddings or topological descriptors may provide additional complementary information while preserving the multiscale philosophy of the framework. Finally, although ABIDE represents one of the largest publicly available autism imaging resources, substantial heterogeneity remains across acquisition sites, participant demographics, and imaging protocols. Rather than weakening the conclusions, this variability emphasizes the need for statistical methods that explicitly address heterogeneous study populations.

Overall, the results consistently support the central premise of this work. 
Functional neuroimaging can be represented at multiple spatial scales, and these representations contribute differently to individualized probability estimation.
Fine-scale connectivity provides detailed discriminatory information, whereas broader network organization improves the stability and reliability of estimated probabilities when prediction models are transferred across heterogeneous imaging environments. Integrating these complementary sources of information therefore yields a richer statistical description of disease than either representation can provide individually.

Although motivated by autism neuroimaging, the framework applies more broadly whenever high-dimensional, spatially dependent measurements are observed at multiple resolutions across heterogeneous institutions, including other biomedical imaging studies, spatial omics, digital pathology, and environmental health research.

More broadly, this work illustrates an important direction for future methodological development in biomedical statistics. As increasingly complex biomedical data become available through large collaborative studies, the challenge will not simply be to develop more sophisticated prediction algorithms. Rather, it will be to construct statistical models that preserve the scientific structure of the data, appropriately quantify uncertainty, and produce probability estimates that remain interpretable and reliable across diverse populations and acquisition environments. The Hybrid Spatial Statistical Learning Framework represents one step toward this objective by combining principles from spatial statistics, statistical learning, and neuroimaging within a unified probabilistic framework motivated by an important biomedical problem.

\section{Conclusion}
\label{sec:conclusion}
The proposed framework demonstrates that complementary spatial representations can improve the quality and transportability of individualized probability estimation without sacrificing discrimination. In the ABIDE application, detailed connectivity supplied most of the participant ranking, whereas broader representations reduced probability error across 18 of 20 held-out sites. This separation of roles is especially relevant for heterogeneous biomedical data, where the representation that discriminates best may not produce the most reliable probabilities.

By bringing spatial structure, site shift, and probability quality into one transparent framework, the method provides a statistically principled foundation for predictive modeling of complex biomedical imaging data collected across heterogeneous institutions.

\printbibliography
\end{document}